\documentclass[11pt]{article}
\usepackage[T1]{fontenc}
\usepackage[utf8]{inputenc}
\usepackage{caption}
\usepackage{subcaption}

\usepackage[english]{babel}
\usepackage{amssymb}
\usepackage{amsmath}
\usepackage{multirow}
\usepackage{lineno}
\usepackage{natbib}
\usepackage{a4wide}
\usepackage{color}
\usepackage{soul}

\usepackage{pstricks}
\usepackage{graphicx}

\newcommand{\Keywords}[1]{\par\noindent
{\small{\em Keywords\/}: #1}}
\usepackage{abstract}

\usepackage{authblk}

\newcommand{\ben}{\vspace{0mm}\begin{equation}}
\newcommand{\een}{\vspace{0mm}\end{equation}}
\newcommand{\be}{\vspace{0mm}\begin{equation*}}
\newcommand{\ee}{\vspace{0mm}\end{equation*}}
\newcommand{\ba}{\vspace{0mm}\begin{equation*}\begin{aligned}}
\newcommand{\ea}{\vspace{0mm}\end{aligned}\end{equation*}}
\newcommand{\ban}{\vspace{0mm}\begin{equation}\begin{aligned}}
\newcommand{\ean}{\vspace{0mm}\end{aligned}\end{equation}}

\usepackage{amsthm}

\begin{document}

\title{Estimation of species relative abundances and habitat preferences using opportunistic data}
\author[1]{Camille Coron \thanks{camille.coron@math.u-psud.fr, corresponding author}}
\author[2]{Cl\'{e}ment Calenge}
\author[1]{Christophe Giraud}
\author[3]{Romain Julliard}
\affil[1]{Laboratoire de Mathématiques d'Orsay, Univ. Paris-Sud, CNRS, Université Paris-Saclay, 91405 Orsay, France.}
\affil[2]{Office national de la chasse et de la faune sauvage, Saint Benoist, BP 20. 78612 Le Perray en Yvelines, France.}
\affil[3]{CESCO, UMR CNRS 7204, Mus\'eum National d'Histoire Naturelle, 55 rue Buffon, 75005 Paris, France}

\setlength{\parindent}{0pt}

\maketitle


\begin{abstract}
We develop a new statistical procedure to monitor, with opportunist data, relative species abundances and their respective preferences for different habitat types. Following \cite{GiraudCalengeCoronJulliard2015}, we combine the opportunistic data with some standardized data in order to correct the bias inherent to the opportunistic data collection. Our main contributions are (i) to tackle the bias induced by habitat selection behaviors, (ii) to handle data where the habitat type associated to each observation is unknown, (iii) to estimate probabilities of selection of habitat for the species. As an illustration, we estimate common bird species habitat preferences and abundances in the region of Aquitaine (France).
\end{abstract}

\section{Introduction}
Citizen science programs have been increasingly developed for biodiversity
monitoring during the last 20 years. These programs usually enroll a 
large number of volunteers to work on a given scientific
issue. For example, breeding bird surveys aim at estimating population
trends of bird species in a given area (\cite{Link1998}); the bird
observations by the volunteers of the program populate a database
describing the number of individuals of every focus species observed
at a given time and place. Since the observational effort is usually much larger 
in citizen science programs than in ``professional''
scientific programs, citizen science programs usually gather
much more observations than classical programs.

The issues tackled by citizen science programs can be very diverse,
including the estimation of the spatial distribution of a set of
species at different spatial scales (\cite{Royleetal2005,Fithianetal2014,GiraudCalengeCoronJulliard2015}), the
study of certain ecological behaviors such as habitat selection (e.g.,
\cite{Biggs2011}), or the monitoring of population trends of
endangered species (\cite{Link1998}). Although some citizen science
programs rely on data collected with standardized protocol and
sampling design (e.g. the North American Breeding Bird Survey,
\cite{Link1998}),
many others rely on the opportunistic collection of observations by
the volunteers, with an unknown observation intensity. 
In the following, we will refer to this
sort of uncontrolled data collection by  ``opportunistic data collection''.
In this paper,
we focus on the estimation of some relative abundances  
based on such opportunistic data.

The opportunistic nature of these data raises important statistical issues (\cite{Dickinson2010}, \cite{Isaacetal2014}). A major issue is due to the non-uniform observation intensity: the collected data cannot be considered as an unbiased sample of the individuals present on this area.  Any statistical approach relying on such
data must tackle this data collection bias in some way.  Some recent
papers (\cite{GiraudCalengeCoronJulliard2015, Fithianetal2014}) proposed to handle this
bias by combining this  biased opportunistic dataset with a
(possibly much smaller) dataset collected in the same area by a more classical
program with a known observational effort (hereafter called
``standardized dataset''). Under
some restrictive assumptions (discussed below), such a combination provides some unbiased estimates of the
relative abundance for the species monitored in at least one of the two programs. 
An attractive feature of these estimation schemes is to provide relative abundance
estimation for species monitored in the opportunistic dataset, but not in the 
standardized dataset. This allows, in principle, to monitor with opportunistic data collection
some rare species that would be much more costly to monitor with 
a classical standardized program.

The approach proposed in  \cite{GiraudCalengeCoronJulliard2015} 
provides, for a set of species, some relative abundance estimates in a collection of sites. 
The statistical modeling accounts for unequal and unknown detectability and reporting rates for  
the monitored species, both in the opportunistic and the standardized dataset, 
and for the unequal  and unknown observational intensity in the opportunistic dataset.
Yet, a crucial hypothesis is that the animals are distributed uniformly 
within each site. When a site gathers several areas with different habitat,  and if the proportions of these different habitats differ among sites,
this assumption is likely to be violated due to habitat selection behavior. 
Similarly the observational effort in opportunistic data is not equally distributed across the different
habitat types due to observer preference for some habitats (\cite{Tulloch2012}). 
This lack of homogeneity induces some important bias in the estimation,  
as shown in \cite{Bellamy1998,MasonMcDonald2004,Fuller2005, Fithianetal2014}.
For example, if the volunteers
participating to a bird monitoring program are mostly interested in
waterbirds, they will strongly select for humid habitat within each
 site. If humid habitat is rare yet present within a site, 
 most of the observations in this site will be performed in this rare habitat, and the
 resulting  waterbird abundance in this site will be
strongly overestimated.

The aim of our paper is to extend  the approach of
\cite{GiraudCalengeCoronJulliard2015} by handling (unknown) habitat
preferences that might influence both observers and observed animal
behaviors. The whole monitored area is described by several
habitat categories for which both observers and animals have different
preferences. The habitat type associated to each observation is \underline{not}
assumed to be known exactly (e.g. the exact location of the observation is only known approximately, or an observed species observation may not be attributed unambiguously to a surrounding habitat). It can be seen as a hidden variable.
Preferences of the observers and of each species 
for each habitat types are also unknown. Our approach provides estimation
for all of them.  Hence, by taking the habitat stratification into account,
we  produce (i) some more accurate relative abundance estimates; in particular, for given site, it allows to decompose a species relative abundance in a habitat-specific component (e.g., forest birds are relatively more abundant because forests are over-represented in that site) and an additive site-specific component, (ii) relative abundance maps at a finer spatial scale, 
and (iii) some estimates of the resource selection functions of the species
(\cite{Manly2002}), which has major implications for
biological conservation.
To sum-up, our main contributions are:
\begin{itemize}
\item To incorporate habitat type preferences in the statistical modeling of \cite{GiraudCalengeCoronJulliard2015};
\item To handle data where the habitat type associated to each observation is unknown (which allows to gather data at different spatial scale);
\item To estimate the relative probabilities of selection of habitat for the monitored species.
\end{itemize}

We  develop our statistical modeling in Section \ref{sectModel}.
 In this new model, the
respective habitat selection behaviors of observers and animals are
modeled using hidden variables. The spatial distribution of observers
in the sites, as well as the habitat selection within the sites is
modeled differently for the two datasets (opportunistic and
standardized). On the other hand, animals are assumed to distribute
within a site according to their preferences for different habitat
types. Then, we illustrate our approach using simulated data to
demonstrate that it recovers the parameters of the model that was used
to simulate the data. Finally, using a real dataset collected on birds
in the Aquitaine region (France), we assess the performance of the
model for estimating species relative abundances as well as their
habitat selection parameters.

\section{Model and parameters}\label{sectModel}

In this section, we introduce our statistical modeling of available data. These data are the outcome of some ecological features (species abundances) and some observational bias (detectability, partial reporting, heterogeneous observational effort, etc). Both the ecological features and the observational bias are affected by some ecological variables (for example habitat type, population and/or road density, altitude, as presented in \cite{MairRuete2016}), which will be called habitats, from now on. Our modeling takes into account this double source of bias induced by the habitat. 
We first describe the ecological ingredients, which are independent from the considered datasets, and then the observational ingredients which are dataset dependent. 

\paragraph{Species abundances and habitat selection probability}
The space-time is divided into units, we call henceforth \emph{sites}, which correspond to the scale at which 
we will predict the relative abundances. So, each site refers to the couple of a spatial domain and a time interval.  We index the sites by $j\in[\![1,J]\!]$. The species we focus on, are indexed by $i\in[\![1,I]\!]$, and we denote by $N_{ij}$ the number of individuals of species $i$ in the site $j$.
Our aim is to estimate the relative abundances $N_{ij}/N_{i1}$ for all $i$ and $j$.

The habitat types of a given site $j$ are not homogeneous.
Each site $j$ gathers several spatial domains, each with a specific habitat type. We index by $h\in [\![1,H]\!]$ the  habitat types. The species $i$ are not uniformly distributed in the spatial domain of $j$: The species $i$ prefer some habitat types to some others and hence are more or less frequent  in the different habitat types. In order to avoid biases in our estimation, we must take this heterogeneity into account. Our modeling assumes that the fraction of the animals of the species $i$ present in the habitat type $h$ inside the site $j$ is proportional to the  \emph{known} area $V_{hj}$ of the habitat type $h$ inside the site $j$ weighted by a number $S_{ih}\in[0,1]$ which represents the preference toward the habitat type $h$ for the species $i$. More precisely, we assume that the density of the species $i$ at location $x$ in the site $j$ is given by 
$$\frac{N_{ij}S_{ih(x)}}{\sum_{h'}S_{ih'}V_{h'j}}\,,$$
with $h(x)$ the habitat type at location $x$. Following the concept definitions clarified in \cite{Lele2013}, the parameters $S_{ih}$ can be interprated as the probability of selection of habitat $h$ by species $i$. These probabilities of selection of habitat are \emph{unknown} and we will estimate them.

\paragraph{Observations and reporting}
As in \cite{GiraudCalengeCoronJulliard2015}, our relative abundance estimation is based on two datasets : (i) a standardized dataset, labeled by $k=0$, collected under a program with a known sampling effort and (ii) an opportunistic dataset, labeled by $k=1$, characterized by a completely unknown sampling effort.  
The datasets gather counts of animals for all sites $j$. We emphasize that each site $j$ must be surveyed by both datasets, and each species $i$ must be surveyed by at least one of the two datasets (at least one species must be surveyed in both datasets). 

We assume that we have informations about the locations of the observations at a finer scale than the site $j$. Each site $j$ is divided into several (possibly many) cells indexed by $c$ and for each observation, we have the information in which cell $c$ the observation occurred. We emphasize that the cell paving can completely differ between the two datasets.  In each dataset, only a (possibly very small) fraction of the cells  have been visited at least once by the observers, so we do \underline{not} have counts for all cells $c$, but only for a (possibly very small) fraction of them. For a cell $c$ visited in the dataset $k$, we denote by $X_{ick}$ the corresponding count for the species $i$. This count $X_{ick}$ is not homogeneously proportional to the abundance of the species $i$ in $c$. Actually, the counts are biased by the inhomogeneous observational effort (total amount of observation time, number of observers, number or density of traps, etc) and the unequal probability of reporting of the species $i$ (varying detectability, partial reporting, etc). Following \cite{GiraudCalengeCoronJulliard2015}, we denote by $\mathcal E_{ck}$ the observation intensity (or effort) in the cell $c$ for the dataset $k$, and by $P_{ik}$ the probability of detection/reporting of the species $i$ in the dataset $k$. When the species $i$ are not monitored in the dataset $k$, the probability of detection/reporting $P_{ik}$ is set to 0.
The model of  \cite{GiraudCalengeCoronJulliard2015} does not take into account habitat types and reads $ X_{ick}\sim\mathcal{P}{\rm oisson}(N_{ij}\mathcal E_{ck}P_{ik})$. For the sake of comparison with the models described below, we scale the effort $\mathcal E_{ck}$ by the area $V_{j}$ of the site $j$ by introducing $E_{ck}=\mathcal E_{ck}V_{j}$. In terms of this scaled effort, the model of   \cite{GiraudCalengeCoronJulliard2015} is
\begin{equation} X_{ick}\sim\mathcal{P}{\rm oisson}(N_{ij}E_{ck}P_{ik}/V_{j}). \label{ModeleNeutre}\end{equation}
Within a cell $c$, the observers do not scan the space uniformly. Actually, they have some preferences for some habitat types (which are not the same for the two datasets). These preferences induce some specific biases, which must be properly addressed. Similarly as for the probability of selection of habitat, the preference of the observers of the dataset $k$ for the habitat $h$ is represented by a real number $q_{hk}\in[0,1]$.
For the dataset $k$, 
we model the observation intensity at location $x$ within the cell $c$ by
$$\frac{q_{h(x)k}E_{ck}}{\sum_{h'}q_{h'k}V_{h'c}},$$ where $V_{hc}$ is the known area of cell $c$ covered by habitat $h$.
Writing  $\mathcal{A}_c$ for the spatial domain of the cell $c$ and taking into account both the probabilities of selection of habitat and the observers habitat preferences, we obtain the modeling for the count of the species $i$ in the cell $c$ for the dataset $k$ 
\ban \label{Model} X_{ick}&\sim \mathcal{P}{\rm oisson}\left(\int_{\mathcal{A}_c}N_{ij}\frac{S_{ih(x)}}{\sum_{h'}S_{ih'}V_{h'j}}\times E_{ck}\frac{q_{h(x)k}}{\sum_{h'}q_{h'k}V_{h'c}}\times P_{ik}\,dx\right)\\&=\mathcal{P}{\rm oisson}\left(N_{ij}E_{ck}P_{ik}\sum_{h} \frac{q_{hk}}{\sum_{h'}q_{h'k}V_{h'c}}\times \frac{S_{ih}}{\sum_{h'}S_{ih'}V_{h'j}}V_{hc}\right).\ean
In the above model, recall that the volumes $V_{hj}$ and $V_{hc}$ are known. For the standardized dataset, the observation intensities $E_{c0}$  are assumed to be known (up to a common multiplicative constant), and we assume that (i) either the habitat type associated to each observation $X_{jc0}$ is known, (ii) or the ratios $q_{h0}/q_{10}$ are known for all $h$ (generally equal to $1$).
All the other parameters are unknown. Their identifiability and the implementation of model \eqref{Model} are detailed in Appendix \ref{sectIdentifiability}. 

We point out that, here, we do not take into account a dependence of the detectability with habitat types (due notably to different levels of visibility in different types of habitat). We refer to Section \ref{sectHabitat} for an extension of this model, 
integrating a dependence of detection probability with habitat types.

Note finally that when neglecting differences in habitat selection probabilities both for observers and observed individuals, the total number $X_{ick}$ of observations of individuals of species $i$ in cell $c$ of domain $j$ for the dataset $k$ follows the model (\ref{ModeleNeutre}) issued from \cite{GiraudCalengeCoronJulliard2015}.

\section{Numerical result}\label{sectResults}

We test our modeling framework both with some simulated data and with some real data.
The likelihood of (\ref{Model}) cannot be maximized easily, so we opt for a non-informative bayesian estimation
 computed with  the Gibbs Sampler \emph{JAGS} (\cite{Plummer2003}). This program is called within \emph{R} (\cite{RCoreTeam2014}) using the \emph{rjags} package (\cite{Plummer2014}. We choose uninformative priors for the unknown parameters and the sampler \emph{JAGS} provides samples distributed according to the posterior distributions for these parameters. The details about the implementation of the estimation procedures are given in Appendix \ref{sectIdentifiability}.

\subsection{Illustration with simulated data}\label{sectSimulateddata}

We  illustrate the ability of our estimation procedure to recover the actual parameter values with some simulated datasets. We compare the results of our procedure to those computed according to the model of \cite{GiraudCalengeCoronJulliard2015}, in order to check whether the gain in model specification counterbalances favorably the inflation of the number of parameters.

We simulate two datasets according to the Model \eqref{Model}: A standardized one ($k=0$) with known relative effort intensities $E_{j0}/E_{10}$ and an opportunistic one ($k=1$) with unknown relative effort intensities. {For this simulated dataset we consider 20 different species on 30 different sites that are covered by $2$ types of habitat. For standardized (resp. opportunistic) data, $10$ (resp. $30$) cells are visited in each site.} The other parameters, such as species abundances, detection probabilities, habitat selection probabilities, or  {efforts in the opportunistic dataset} are sampled according to {uniform distributions}. 

In order to illustrate the impact of the habitat modeling and the gain of using opportunistic data, we compare 
 the three following estimation procedures:\medskip

\begin{tabular}{c p{0.7\textwidth}}
$[$Opp+Stand with hab$]$ & Our model  \eqref{Model} with unknown habitat selection probabilities and using both opportunistic  and standardized  data, denoted below by [Opp+Stand with hab]; \medskip\tabularnewline 
$[$Stand only with hab$]$ & Our model \eqref{Model} with unknown habitat selection probabilities and using \underline{only} standardized data (which corresponds to Equation \eqref{Model}, with $k=0$ only). It is denoted by [Stand only with hab]; \medskip\tabularnewline
$[$Opp+Stand no hab$]$ & The model \eqref{ModeleNeutre} 
which neglects differences in selection probabilities, using both opportunistic and standardized data. This model is denoted by [Opp+Stand no hab].
\end{tabular}
\medskip

\begin{figure}[!ht]
\begin{center}
    \includegraphics[scale=0.44]{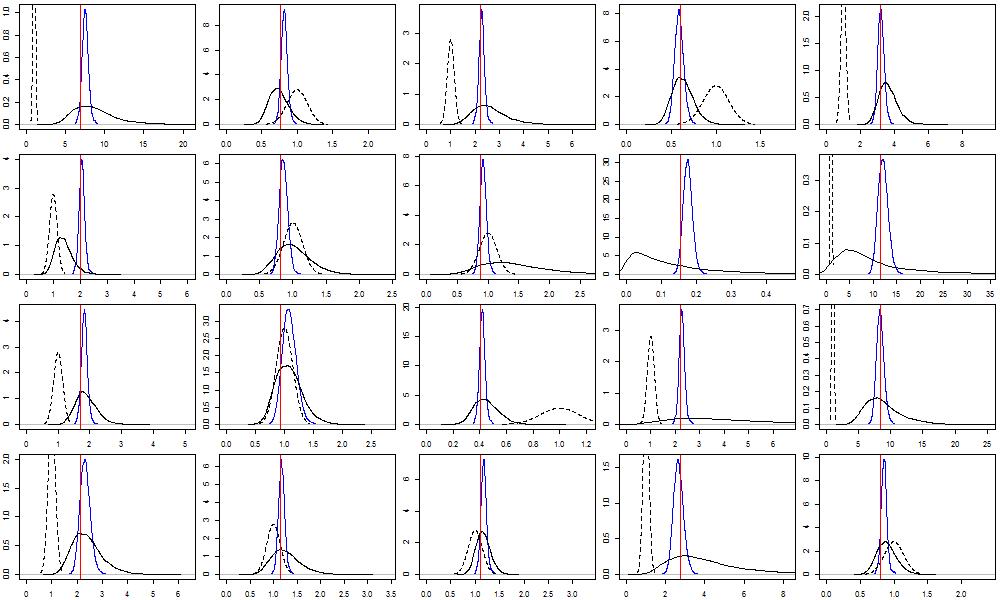}
    \caption{Posterior distributions of the relative abundances $N_{i2}/N_{i1}$ for all $i$,  estimated with [Stand only with hab] (in black), [Opp+Stand with hab] (in blue) and [Opp+Stand no hab] (dotted line). The reference values are given in red.}
    \label{FigureDonneesSimuleesAbondances}
\end{center}
\end{figure}

\begin{figure}[ht]
\begin{center}
\includegraphics[scale=0.25, trim=0cm 0cm 0cm 0cm]{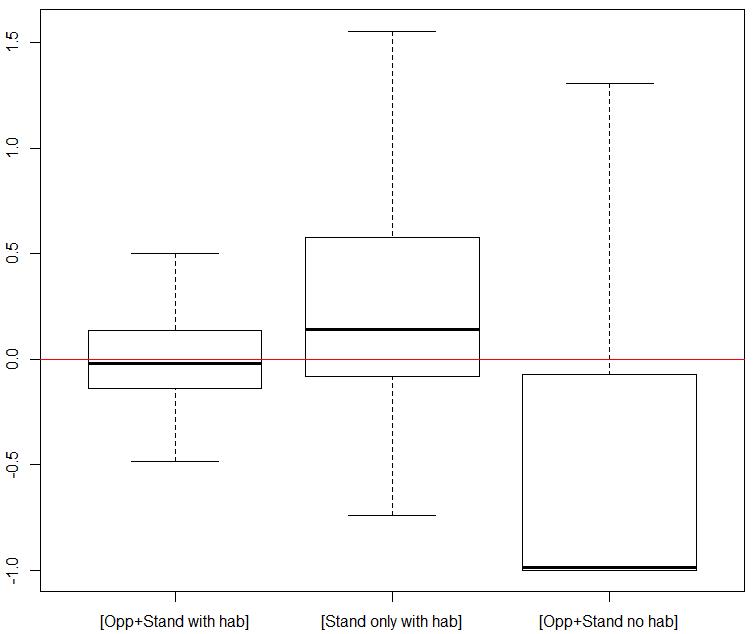}
    \caption{Boxplots of the relative differences {$\frac{({\hat{N}^{[model]}_{ij}}/{\hat{N}^{[model]}_{i1}})-({N_{ij}}/{N_{i1}})}{{N_{ij}}/{N_{i1}}}$ between the estimated and "real" relative abundances, using data and models [Opp+Stand with hab], [Stand only with hab], and [Opp+Stand no hab]}.
}
    \label{FigureBoxplotsRelDiffSD}
\end{center}
\end{figure}

 In Figure \ref{FigureDonneesSimuleesAbondances}, we plot the posterior distributions of relative species abundances obtained for these three models, and the  reference relative species abundance values that we  estimate are given in red. This figure shows, as proved in \cite{GiraudCalengeCoronJulliard2015}, the improvements brought by opportunistic data, since the estimation obtained by combining the two datasets is both more precise and more accurate. It also illustrates that neglecting habitat preferences can lead to biased estimation of species relative abundances. Figure \ref{FigureDonneesSimuleesSpecificites} in Appendix \ref{AppResultsSD} also gives the posterior distributions of habitat selection probabilities, that give good approximations of the real values given in red. Figure \ref{FigureBoxplotsRelDiffSD} gives the boxplots of the relative differences between the estimated and real relative abundances, for each of the three models. Here, the estimated relative abundances is defined as the mean of the associated posterior distribution, and similar results are obtained using the median of these distributions. Again, we observe that the estimation combining both datasets and taking habitat types into account produces better results; and ignoring habitat types induces a significant bias.

\subsection{Real data}\label{sectRealdata}

\subsubsection{Datasets and Habitats}\label{subsectdataset}
\paragraph{Datasets}

To investigate whether taking the habitat types into account improves the estimation on real data, we consider the same datasets as in \cite{GiraudCalengeCoronJulliard2015}. These are two different datasets of common birds observations in Aquitaine (south-western French region): standardized data are provided by the French National Hunting and Wildlife Agency (ONCFS, Office National de la Chasse et de la
Faune Sauvage), by the French National Hunters' Association (FNC,
F\'{e}d\'{e}ration Nationale des Chasseurs) and by the French
Departemental Hunters' Associations (FDC, F\'{e}d\'{e}rations
D\'{e}partementales des Chasseurs), while opportunistic data are provided by the French program Faune-Aquitaine, managed by the protection association Ligue pour la Protection des Oiseaux (LPO). Our estimation is assessed using a validation dataset, produced by the French Museum of Natural History.

\begin{figure}[ht]
\begin{center}
\includegraphics[scale=0.35, trim=0cm 5cm 0cm 4cm, clip]{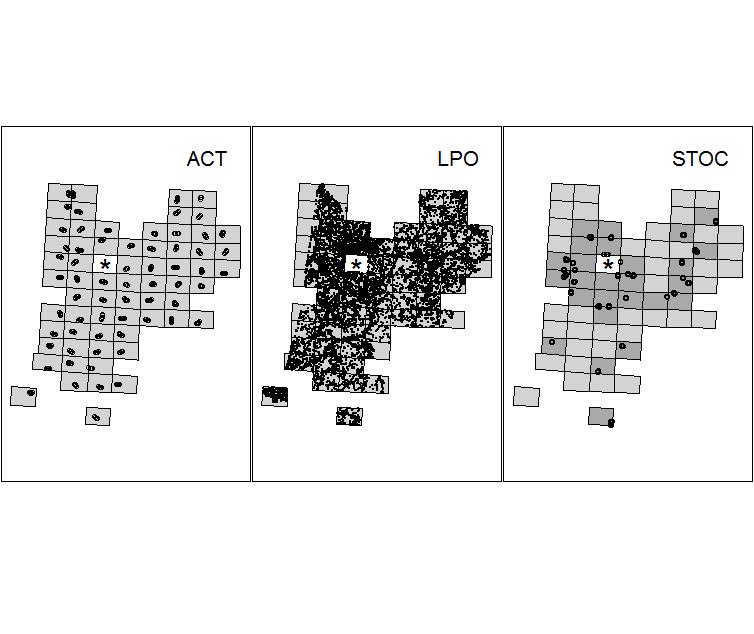}
    \caption{Positions of data collecting}
    \label{FigureDonnees}
\end{center}
\end{figure}

For the first (standardized) dataset we used the ACT monitoring survey (see \cite{Boutin2003} for more details concerning this dataset and its protocole) in which $13$ species of birds are monitored (see Table \ref{TableSpecies}). The observers are
professionals from the technical staff of the participating organisms. The Aquitaine region was
discretized into 66 quadrats, in which a 4-km-long route
was randomly placed in non-urban habitat (see
Figure \ref{FigureDonnees}). Each route was traveled twice between April and
mid-June and included 5 points separated by
exactly 1 km: at each travel, each point was visited for exactly 10 minutes. The species of every bird
heard or seen was recorded, and for each point and each species, we have access to the
maximum of the counts from the two visits (in order to take advantage of the maximum detectability and to avoid effects due to migration, as explained in \cite{Pollock1982}). This protocol was repeated for several years and we use data from 2008 to 2011, which finally leads to 239 visits of quadrats (some of the quadrats were not visited each year), therefore leading to $13*5*239=15535$ data, corresponding to the reporting of $7899$ birds observations (some species are not always detected).

For opportunistic data, we used the dataset collected by the website www.fauneaquitaine.org (handled by the LPO), on which anyone can register and report the species, number of detected individuals, date, and location associated to any bird observations made in Aquitaine. The level of precision of the location is variable: exact location, locality indication, or commune indication. To deal with this inhomogeneity in location information, for numerical analyses we will use the commune in which was made each observation, which is always given. As previously, we selected all such records between April and mid-June for the years 2008--2011. This led to $693~581$ birds observations in $1622$ communes (see Figure \ref{FigureDonnees}), monitoring $34$ species. Note that, to make their observations, observers can go anywhere, for an unknown amount of time, and that they report their observations with an unknown probability (that might depend on the observed species); therefore these data do not provide any information concerning observation effort.

For the validation dataset used to assess the
predictive power of our approach, we used the data from the STOC program
(\textit{Suivi temporel des oiseaux communs}), which is a French breeding bird
survey carried out by the French Museum of Natural History (MNHN,
Museum National d'Histoire Naturelle). The protocole of this survey (see \cite{Jiguetetal2012} for more details) is the following: each observer is assigned a $2\times2$ km square whose position is uniformly randomly chosen within $10$ km of his/her house. The observer then distributes on the considered square, $10$ observation points that have to be representative of the different habitats areas on the square, and each point is visited twice between April and mid-June, during $5$ minutes. Every observation of each species (hearing or seeing) is reported and the maximum count among the two visits is kept, as for the ACT program. As previously, we use all such records for the years 2008--2011. This leads to $86526$ birds observations in $38$ squares (see Figure \ref{FigureDonnees}), monitoring $34$ species (the same than for the LPO dataset).

\paragraph{Habitats}

Land use (habitat) types were based on corine landcover typologies that were grouped in 7 categories in order to reduce complexity and ensure identifiability (see Appendix \ref{sectIdentifiability}): urbanized area, intensive agriculture with homogenous landscape (arable land or permanent crop), open natural landscape (natural or pasture), farmland with heterogenous landscape, mixed forest, deciduous forest, and coniferous forest.

\subsubsection{Assessing estimation performances}
In this section, we compare the estimation performances of the same three  statistical models as in the Section \ref{sectSimulateddata}: [Opp+Stand with hab], [Stand only with hab] and [Opp+Stand no hab]. 
We obtain estimation for the main parameters of interest of the model, namely the relative abundances $N_{ij}/N_{i1}$ for all $i,j$, and the habitat selection probabilities $S_{ih}/S_{i1}$ for all $i,h$ and $q_{h2}/q_{12}$ for all $h$. 
Our goal here is to compare the three estimation schemes, so we do not to discuss the ecological aspects of our estimates.
Yet, in Appendix \ref{AppResultsRD}, we provide some 
 abundances maps and some habitat selection probabilities for some species of interest.

To assess the performances of the three models, we investigate their ability to predict the STOC observations $X_{ij}^{STOC}$, in each quadrat surveyed in the STOC dataset. 
Since some species {(21 among 34, see Appendix \ref{AppList} for the exact list)} are not surveyed in the ACT dataset, we split apart the results for the species surveyed in ACT and the results for the others. 
For each species $i$ and each of the three models, we compute a predictor $\widehat X_{ij}^{model}$ (described in Appendix  \ref{AppNumerics}) of $X_{ij}^{STOC}$. Then for each species $i$ and each model we compute the Pearson correlation between the vector $(\widehat{X}^{model}_{ij})_{j}$ and the vector $(X^{STOC}_{ij})_{j}$.
 The medians (as well as the first and third quartiles) of these correlations (calculated for each species $i$) are given in Table \ref{TableCorrelation}. We notice that the results for the  Model \eqref{ModeleNeutre} slightly differ from the results from \cite{GiraudCalengeCoronJulliard2015}, this can be explained by the fact that we use non-informative Bayesian estimation performed with JAGS instead of maximum likelihood estimation.
 \begin{table}[ht]
\begin{center}
\begin{small}
\begin{tabular}{l|cc}
\hline 
Data and model & Correlations (In ACT) & Correlations (not in ACT)\\
\hline
{[Opp+Stand with hab]} &0.49 (0.30--0.54) & 0.39 (0.12--0.54)\\
{[Stand only with hab]} & 0.29 (0.03--0.46) & --\\  
{[Opp+Stand no hab]} & 0.44 (0.32--0.68) & 0.31 (0.19--0.42) \\
\hline
\end{tabular}
\caption{Medians of Pearson correlation coefficients {(as well as first and third quartiles)} between the STOC observations and the estimates of species relative abundances {computed with the models [Opp+Stand with hab], [Stand only with hab], and [Opp+Stand no hab].}}
\label{TableCorrelation}
\end{small}
\end{center}
\end{table}

\begin{figure}[ht]
\begin{center}
\includegraphics[scale=0.3, trim=0cm 0cm 0cm 0cm, clip]{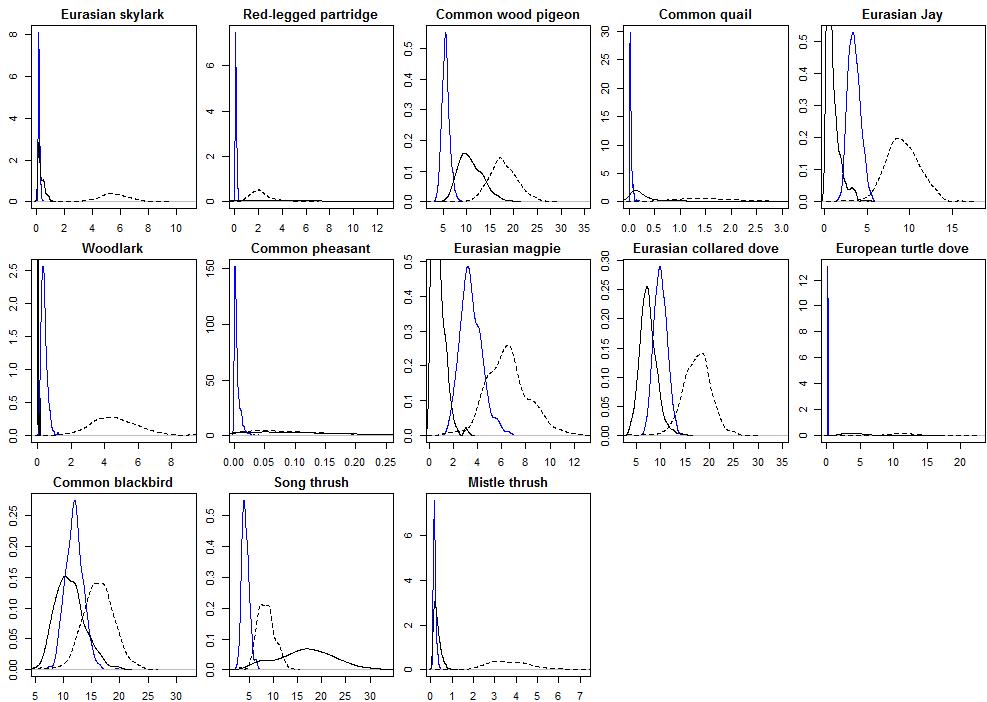}
    \caption{{Posterior of the relative abundances $\hat{N}^{[model]}_{ij}/\hat{N}^{[model]}_{i1}$ of species $i$ monitored in the ACT program and $j=62$, and using data and models [Opp+Stand with hab] (blue), [Stand only with hab] (black), and [Opp+Stand no hab] (dotted line).}}
    \label{FigurePosteriors}
\end{center}
\end{figure}

Figure \ref{FigurePosteriors} shows the posterior distributions of the relative abundances $N^{[model]}_{ij}/N^{[model]}_{i1}$ obtained for the three considered situations.

We observe an improvement of the predictions when we take the habitat into account. Actually, when we move from the Model \eqref{ModeleNeutre} without habitat modeling, to the Model \eqref{Model} with habitat modeling, the median of the correlations between the vectors $(\widehat{X}^{model}_{ij})_{j}$ and $(X^{STOC}_{ij})_{j}$ increases from 0.44 to 0.49 for the species surveyed in the standardized dataset (ACT) and from 0.31 to 0.39 for the other species. So, the reduction of the bias obtained by modeling the habitat is stronger than the increase of the variance induced by the inflation of the number of parameters. This feature is confirmed by the difference between the BIC value for {the model [Opp+Stand with hab] and the model [Opp+Stand no hab]}: 
$$\Delta BIC=BIC({\rm Opp+Stand\ with\ hab})-BIC({\rm Opp+Stand\ no\ hab})=-8867.$$
This improvement can be explained by the strong differences in habitat selection parameters $S_{ih}$ that we have estimated (see Figure \ref{FigurePreferences}), confirming that habitat biases cannot be ignored. This feature is also supported by the shape of the posterior distribution of the model with habitat [Opp+Stand with hab] which is much more spiked than the shape of the posterior distribution of  the model without habitat [Opp+Stand no hab].

In addition, as in \cite{GiraudCalengeCoronJulliard2015}, we observe that adding the opportunist data in our estimation 
improves the estimation. In particular, we observe that for our dataset, the improvement brought from the use of opportunistic data is stronger than the improvement brought from habitat modeling. In the next paragraph, we investigate the importance of the habitat structure in the spatial variation of abundance.

\subsubsection{Models of spatial repartition}
In our model \eqref{Model}, we incorporate two sources of abundance variation. Part of the  variation in abundance is explained by the habitat structure. This part is driven by the habitat selection probability $S_{ih}$. The remaining of the abundance variation comes from some other factors, acting differentially on the different sites $j$.
We investigate below whether the spatial repartition of individuals is mainly explained by habitat structure, or if some other factors have a major role. In the first case, most of the variation in the spatial variation would be explained by the variations of the ratio ${S_{ih(x)}}/\left({\sum_{h}S_{ih}V_{hj}}\right)$; while in the second case, a major part of the spatial variation would be explained by the variations with $j$ of the  $N_{ij}$. In order to compare the relative importance of both effects, we compare the models derived from \eqref{Model} by first neglecting the variation in the habitat selection probabilities $S_{ih}$ (setting them all  to 1); second by neglecting the variation in $j$ of the $N_{ij}$, replacing all of them by a single value $N_{i}$. As explained before, when all the  habitat selection probabilities $S_{ih}$ are equal, the model \eqref{Model} reduces to the model  \eqref{ModeleNeutre} of \cite{GiraudCalengeCoronJulliard2015}.
So to investigate the relative importance of the habitat types with respect to  the other factors, we compare the model [Opp+Stand with hab] and [Opp+Stand no hab] with the [One Quadrat with hab] model where\medskip

\begin{tabular}{c p{0.7\textwidth}}
$[$One Quadrat with hab$]$ & 
$\displaystyle{X_{ick}\sim\mathcal{P}{\rm oisson}\left(N_{i}E_{ck}P_{ik}\sum_{h} \frac{q_{hk}}{\sum_{h'}q_{h'k}V_{h'c}}\frac{S_{ih}}{\sum_{h'}S_{ih'}V_{h'}}V_{hc}\right)}$,
 where $V_h$ is the total area of habitat $h$ in the considered space.
\end{tabular}
\medskip

For each model, we compute as previously the Pearson correlations between the observations of the STOC dataset and the estimation of these observations using each of the three models. The median and the quartiles for all the species are given in Table \ref{TableCorrelationHabitat}.
\begin{table}[ht]
\begin{center}
\begin{small}
\begin{tabular}{l|c}
\hline 
Data and model & Correlations \\
\hline
\text{[Opp+Stand with hab]} &0.45 (0.23--0.54) \\
 \text{[One Quadrat with hab]} & 0.15 (-0.03--0.36) \\  
 \text{[Opp+Stand no hab]} & 0.38 (0.26--0.52) \\
\hline
\end{tabular}
\caption{Medians (as well as first and third quartiles) of Pearson correlation coefficients between estimation of species relative abundances using different models of species spatial repartition.}
\label{TableCorrelationHabitat}
\end{small}
\end{center}
\end{table}

We observe that, in our case, the overall fit for the [one quadrat] model  is poorer. This point is confirmed by the BIC values
$$\Delta BIC = BIC({\rm [Opp+Stand\ no\ hab]})-BIC({\rm [One\ Quadrat\ with\ hab]})=-813853.$$
The habitat type therefore does not seem to be the main driver of the spatial variation for most of the considered species on the considered space (though this ecological result can be different when considering other species on an other space, notably with more diverse types of habitats). The study of the impact of habitat structure on the estimates for each species relative abundance is presented in Section \ref{secDiscussion}. 
 
\section{Discussion}\label{secDiscussion}

\paragraph{Main results.}
We have developed a new statistical approach relying on the joint use
of two datasets collected respectively by an opportunistic data collection program and by a classical
standardized monitoring program with a known (and ideally controlled)
observation intensity. By combining these two datasets, our approach
estimates the relative abundance of a set of species in a set of
sites, while accounting for the different detectability of the species
in the two programs, variable habitat preferences by both the species
and the observers, and unknown observation intensity in the opportunistic data collection program. The use of opportunistic data in this approach
results in a considerably increased precision in comparison to the
estimation that would be based only on the standardized data. Note
also that our approach allows the estimation of the relative abundances
of some species monitored only with the opportunistic data collection program, as
long as there are at least several other species monitored by the two
programs. Our approach extends the statistical modeling developed in
\cite{GiraudCalengeCoronJulliard2015}, by taking into account 
the variable preferences of habitat types by both the
species and the observers. We show in the present paper that by
accounting for habitat preferences by both the species and the
observers in the citizen science program, our approach results in
less bias and an increased performance of the relative abundances estimates. This is illustrated by
a simulated dataset, as well as a practical case study.

A useful byproduct of this approach is the estimation of the
relative preferences of each species for each habitat type: more
precisely, the estimated value of the habitat selection parameters $S_{ih}$ corresponds to the relative
probability of selection of habitat $h$, which is exactly the
definition of a resource selection function (RSF, \cite{Lele2013}), a
tool widely used in biological conservation and wildlife management to
identify important habitat for a given species on a study area
(\cite{Boyce1999}). Existing statistical approaches for the fit of RSF
rely on the comparison of an unbiased sample of the habitat used by
the focus species, and an unbiased sample of either the unused habitat
or the habitat available to the species (see a list of possible
statistical approaches in \cite{Manly2002}, especially chapter 4 for
the case where habitat is defined by several categories). The
collection of such data can be expensive, and when the study area is
large and/or the focus species is rare it can become prohibitive (see the conclusion of \cite{MacKenzie2005a}). However, endangered rare species are precisely
those for which information on selected places is the most
crucial. The situation is generally worsened when several rare and
endangered species are under study. Citizen science programs relying
on opportunistic data collection are a very attractive alternative in
this context because of the large observation effort carried out, but
are often notoriously flawed by an unequal and unknown observation
effort, which make their use in such studies difficult
(\cite{Phillips2009}). If at least a part of the species monitored in
the opportunistic data collection program are also monitored in a more classical
standardized program, our approach provides a way to correct for the
biases caused by the unequal observation effort in opportunistic data collection programs, and therefore to benefit of the large observation effort for
the RSF estimation. Our approach therefore allows the batch estimation
of the RSF for all species in the opportunistic data collection program.

\paragraph{Limitations.}
Our approach relies on the hypothesis that the preferences of a given
species for a given habitat type does not vary into space and
time. Several authors have shown that this might not always be the
case: animals sometimes show a functional response of habitat
selection, i.e. an habitat selection pattern that depends on habitat
availability (\cite{Mysterud1998}); an habitat type can therefore be
preferred by a species in a context and avoided in another
(e.g. \cite{Calenge2005}). Similarly, our approach supposes that the
observers in the citizen science program show a constant preference in
space and time. However, the observers preferences can also be
characterized by functional responses. For example, in an opportunistic data collection program focusing on birds, observers may be more interested by
waterbirds in humid regions and therefore prefer to spend their time
close to lakes and ponds in such regions, as this is where they are
more likely to observer the species of interest. On the other hand, in
a mountainous region, observers might be more interested into raptors
and avoid lake and ponds. Such functional response of the observers
can bias the resulting estimates, and should be seriously considered
when fitting this model.

\paragraph{Possible extension.}
So far, we assumed that the detectability $P_{ik}$ of species $i$ in
dataset $k$ does not depend on the habitat type, which might be
unrealistic since, in particular, the range of vision of an observer
can be different from one habitat type to the other. If so, our
estimation of habitat selection parameters $S_{ih}$  can be biased. 
 Due to identifiability
constraints, we cannot include in the model and estimate an unknown
list of parameters $\alpha_h$ taking into account the dependence of
detectability to habitat (since they will be undistinguishable from
the species habitat selection probabilities $S_{ih}$). However, it is
possible to include these parameters in the model and define an
informative prior distribution on these detectabilities, if
information is available elsewhere. Another solution is to use 
additional data concerning the
detectability associated to each considered habitat. We demonstrate in
Appendix \ref{sectHabitat} how to implement this approach with the dataset
used in this paper, by using additional data that give for
different kinds of habitats the respective numbers of observations
made in different ranges of distances. Based on an idea similar to the
statistical approach underlying the abundance estimation based on
distance sampling (\cite{Buckland1993}), we demonstrate how to
estimate and account for the variable detectability between habitat
types when estimating the species relative abundance in each site.

\paragraph{Issues when implementing the model.}
Several issues must be carefully handled when implementing our model on some datasets. 
A key step in the implementation of our model lies in the choice of the habitat types and their number.
This choice, which is dataset dependent, must be handled with care.
First, the choice of the habitat types must be meaningful for 
the monitored species. For example, assume that some of the monitored species 
have a very different selection probability for 
two given habitat types, say "deciduous forest" and "coniferous forest".
If the proportion of "deciduous forest" and "coniferous forest" 
varies from one site to the other, then the merging  of these two habitat types 
into a single habitat type "forest" would induce a significant bias in the estimation.
To avoid such biases, we may be tempted to select a very large number of 
habitat types, ensuring a strong homogeneity of each type.
Yet, the multiplication of the habitat types is limited by 
the number of available observation points in each habitat type. Actually, 
in order to avoid a detrimental increase of the variance,
we need to have enough observation points in each habitat type.
These observations can be indifferently in the opportunistic or in the standardized dataset.
So, when choosing the habitat types (and their number), one 
must find a good balance between defining meaningful habitat types for the
monitored species and having enough observations in each habitat type.

Another major degree of freedom in the implementation of the model, is
the choice of the number of species. On the one hand, increasing the number of
species helps the estimation, since the ratio between the number of observations 
and the number of parameters then decreases. On the other hand, including
some rare species, or including some species which require to 
add some new habitat types, can harm the estimation by  
increasing the variance. So, again, a good balance must be found between the two phenomena.

For a fair comparison of the statistical models with and without habitat modeling,
we have implemented our model with very flat priors on all the parameters.
In practice, we may have access to some existing estimates for some of the parameters.
For example, for some species, we can have some estimates for some
of the habitat probabilities of selection (\cite{Manly2002}).
In this case, it is worth to incorporate this knowledge by designing
some more informative priors on the habitat probabilities of selection.

{\bf Acknowledgements:} {\sl  The authors would like to thank Benjamin Auder for his contribution to the computer code for data treatment. We sincerely thank Denis Roux and all observers from the ACT network. We also thank the managers of both programs « Faune d’Aquitaine » and STOC-EPS, as well as the observers participating to these programs.  This work  was partially funded by public grants as part of the
"Investissement d'avenir" project, reference ANR-11-LABX-0056-LMH,
LabEx LMH, and reference ANR-10-CAMP-0151-02, Fondation Math\'ematiques Jacques Hadamard, by the Chair "Mod\'elisation Math\'ematique et Biodiversit\'e" of VEOLIA-Ecole Polytechnique-MNHN-F.X, and by the Mission for Interdisciplinarity at CNRS.} 

\appendix

\section{Appendix}

\subsection{List of species}\label{AppList}

Table \ref{TableSpecies} provides the list of the 34 bird species under study. 

\begin{table}
\vspace{-2cm}
\begin{center}
  \caption{List of the 34 bird species under study. The 13 species
    that are monitored by the ACT survey are indicated by an
    asterisk.
  } 
  \label{TableSpecies}
      \begin{scriptsize}
  \begin{tabular}{ll||ll}
    \hline
    Latin name & Species  & Latin name & Species\\ 
    \hline
    \textit{Aegithalos caudatus} & Long-Tailed Tit &
    \textit{Alauda arvensis}$^*$ & Eurasian Skylark  \\
    \textit{Alectoris rufa}$^*$ & Red-Legged Partridge  &
    \textit{Carduelis carduelis} & European Goldfinch  \\
    \textit{Carduelis chloris} & European Greenfinch &
    \textit{Certhia brachydactyla} & Short-Toed Treecreeper \\
    \textit{Columba palumbus}$^*$ &  Common Wood Pigeon  &
    \textit{Coturnix coturnix}$^*$ & Common Quail \\
    \textit{Cuculus canorus} & Common Cuckoo &
    \textit{Cyanistes caeruleus} & Eurasian Blue Tit  \\
    \textit{Dendrocopos major} & Great Spotted Woodpecker &
    \textit{Erithacus rubecula} & European Robin  \\
    \textit{Fringilla coelebs} & Common Chaffinch &
    \textit{Garrulus glandarius}$^*$ & Eurasian Jay \\
    \textit{Hippolais polyglotta} & Melodious Warbler &
    \textit{Lullula arborea}$^*$ & Woodlark  \\
    \textit{Luscinia megarhynchos} & Common Nightingale &
    \textit{Milvus migrans} & Black Kite \\
    \textit{Parus major} & Great Tit  &
    \textit{Passer domesticus} & House Sparrow \\
    \textit{Phasianus colchicus}$^*$ & Common Pheasant &
    \textit{Phoenicurus ochruros} & Black Redstar \\
    \textit{Phylloscopus collybita} &  Common Chiffchaff &
    \textit{Pica pica}$^*$ & Eurasian Magpie  \\
    \textit{Pica viridis} & Eurasian Green Woodpecker  &
    \textit{Sitta europaea} & Eurasian Nuthatch  \\
    \textit{Streptopelia decaocto}$^*$ & Eurasian Collared Dove &
    \textit{Streptopelia turtur}$^*$ & European Turtle Dove \\
    \textit{Sylvia atricapilla} & Eurasian Blackcap &
    \textit{Troglodytes troglodytes} & Eurasian Wren \\
    \textit{Turdus merula}$^*$ & Common Blackbird &
    \textit{Turdus philomelos}$^*$ & Song Thrush \\
    \textit{Turdus viscivorus}$^*$ & Mistle Thrush &
    \textit{Upupa epops} & Eurasian Hoopoe  \\
    \hline
  \end{tabular}
      \end{scriptsize}
  \end{center}
\end{table}

\subsection{Identifiability and models implementation}\label{sectIdentifiability}
\subsubsection{Reparametrization of the model}
Let us recall our model for the observations, where $c$ is a cell of the site $j$:

\ba X_{ick}&\sim \mathcal{P}{\rm oisson}\left(\int_{\mathcal{A}_c}N_{ij}\frac{S_{ih(x)}}{\sum_{h'}S_{ih'}V_{h'j}}\times E_{ck}\frac{q_{h(x)k}}{\sum_{h'}q_{h'k}V_{h'c}}\times P_{ik}\,dx\right)\\&=\mathcal{P}{\rm oisson}\left(N_{ij}E_{ck}P_{ik}\sum_{h} \frac{q_{hk}}{\sum_{h'}q_{h'k}V_{h'c}}\frac{S_{ih}}{\sum_{h'}S_{ih'}V_{h'j}}V_{hc}\right).\ea

For standardized data, we can assume either that $q_{h0}/q_{10}$ is known for all $h$ (generally equal to $1$), or that each cell for the standardized dataset is small enough to be composed with only one habitat. In addition, we assume that $E_{c0}/E_{10}$ is known for all $c$ for the standardized dataset. To implement our model while ensuring identifiability of the  parameters, we use the following change of variables
\be \tilde{N}_{ij}=\frac{N_{ij}P_{i0}E_{10}}{\sum_{h'}\frac{S_{ih'}}{S_{i1}}V_{h'j}},\quad 
\tilde{P}_{ik}=\frac{P_{ik}P_{10}}{P_{i0}P_{1k}}, \quad
\tilde{E}_{ck}=\frac{E_{ck}P_{1k}}{P_{10}E_{10}}\frac{V_c}{\sum_{h'}\frac{q_{h'k}}{q_{1k}}V_{h'c}},\ee
\be
\tilde{q}_{hk}=\frac{q_{hk}}{q_{1k}},\quad
\tilde{S}_{ih}=\frac{S_{ih}}{S_{i1}}
\ee

where $V_c=\sum_hV_{hc}$. Using this change of variables, we get that for all $i$, $c$ and $k$,

$$X_{ick}\sim \mathcal{P}\left(\tilde{N}_{ij}\tilde{E}_{ck}\tilde{P}_{ik}\sum_{h}  \tilde{q}_{hk}\tilde{S}_{ih}V_{hc}\right)$$

with

\be \frac{\tilde{N}_{ij}}{\tilde{N}_{i1}}=\frac{N_{ij}}{N_{i1}}\frac{\sum_{h'}\tilde{S}_{ih'}V_{h'1}}{\sum_{h'}\tilde{S}_{ih'}V_{h'j}},\quad
\tilde{P}_{i0}=1,\quad \tilde{P}_{11}=1,\quad 
\tilde{q_{h0}}=1, \quad \tilde{q_{11}}=1, \quad \tilde{S_{i1}}=1
\ee
for all $i$, $c$, $k$, and $\tilde{E}_{c0}$ is known for all $c$.

In particular for standardized data, for which we can assume that the habitat associated to each cell $c$ is known (denoted by $h(c)$), we get:

$$X_{ic0}\sim \mathcal{P}\left(\tilde{N}_{ij}\tilde{E}_{c0}\tilde{S}_{ih(c)}\right),$$

where $\tilde{E}_{c0}$ is known for all cell $c$. This is a generalized linear model with $IJ+I(H-1)$ unknown parameters (the quantities $\tilde{N}_{ij}$ as well as habitat selection parameters $\tilde{S}_{ih}/S_{i1}$ for $h>1$). These parameters are identifiable if and only if the matrix $Y$ with size $C\times (J+H-1)$ giving for each cell $c$ visited by the STOC dataset, the site and habitat associated to this cell (when this habitat is not the first habitat), has rank $J+H-1$. More precisely, the matrix $Y$ is such that for all $c\in[\![1,C]\!]$, $Y_{cj(c)}=1$, $Y_{c{J+(h(c)-1)}}=1$ if $h(c)>1$, and $Y_{cl}=0$ elsewhere.

\subsubsection{Implementation with JAGS}

{The computer code associated to the section \ref{sectSimulateddata} is given in the numerical Additional File SimulatedData.Rnw.
This program calls three models that are written in separate files: one for our model (Additional file ModelSimulatedData.txt), one for the model in which we use only standardized data (Additional file ModelStandardizedSimulatedData.txt), and one for the model in which differences in habitat preferences are neglected (Additional file ModelWithoutHabitatSimulatedData.txt).}

{The computer code associated to the section \ref{sectModel} is given in the numerical Additional File RealData.Rnw. This program calls four models that are written in separate files: one for our model (Additional file ModelWithHabitat.txt), one for the model in which we use only standardized data (Additional file ModelStandardizedOnly.txt), one for the model in which differences in habitat preferences are neglected (Additional file ModelWithoutHabitat.txt), and one for the model in space is considered as one single quadrat (Additional file ModelOneQuadrat.txt).}

\subsection{Some details on the numerics: the prediction of the STOC data}
\label{AppNumerics}
Let $\mathcal C^{STOC}_{j}$ denote the set of all the observation points  $c$ in the quadrat $j$ surveyed in the STOC dataset.  
The STOC counts for the species $i$ in the quadrat $j$ are
$$X_{ij}^{STOC}=\sum_{c\in \mathcal C^{STOC}_{j}} X_{ic}^{STOC}.$$
Let us denote by $h(c)$ the habitat type of the  observation point $c$.
In our model \eqref{Model}, the average number of individuals of the species $i$ in the square $c\in \mathcal C^{STOC}_{j}$ is given by 
$$\int_{\mathcal{A}_{c}}\frac{N_{ij} S_{ih(c)}}{\sum_{h'} S_{ih'}V_{h'j}} \,dx = \frac{N_{ij} S_{ih(c)}V_{c}}{\sum_{h'} S_{ih'}V_{h'j}}.$$
Taking into account a variable observational effort $E_{c}^{STOC}$ on each observation point $c$, we then predict  $X_{ij}^{STOC}$ from the estimation based on our Model \eqref{Model} by
$$\widehat X_{ij}^{model}=\hat N^{model}_{ij} \sum_{c\in \mathcal C^{STOC}_{j}} E_{c}^{STOC} \frac{\hat S^{model}_{ih(c)}V_{c}}{\sum_{h'} \hat S^{model}_{ih'}V_{h'j}},$$
where the observational effort $E_{c}^{STOC}$ is given by the number of  years of observation at the observation point $c$.

For the one quadrat model with habitat [One Quadrat with hab] displayed in Table \ref{TableCorrelationHabitat}, the prediction is given by
$$\widehat X_{ij}^{model}=\hat N^{model}_{i} \sum_{c\in \mathcal C^{STOC}_{j}} E_{c}^{STOC} \frac{\hat S^{model}_{ih(c)}V_{c}}{\sum_{h'} \hat S^{model}_{ih'}V_{h'}},$$
with $V_{h'}$ the area of the habitat type $h'$ in the whole quadrat.

When the Model \eqref{ModeleNeutre} is used for estimation, then the predictions are given by 
$$\widehat X_{ij}^{model}=\hat N_{ij} \sum_{c\in \mathcal C^{STOC}_{j}} E_{c}^{STOC}\frac{V_{c}}{V_{j}}.$$

\subsection{Additional results on simulated data}
\label{AppResultsSD}
\begin{figure}
\begin{center}
   \includegraphics[scale=0.3]{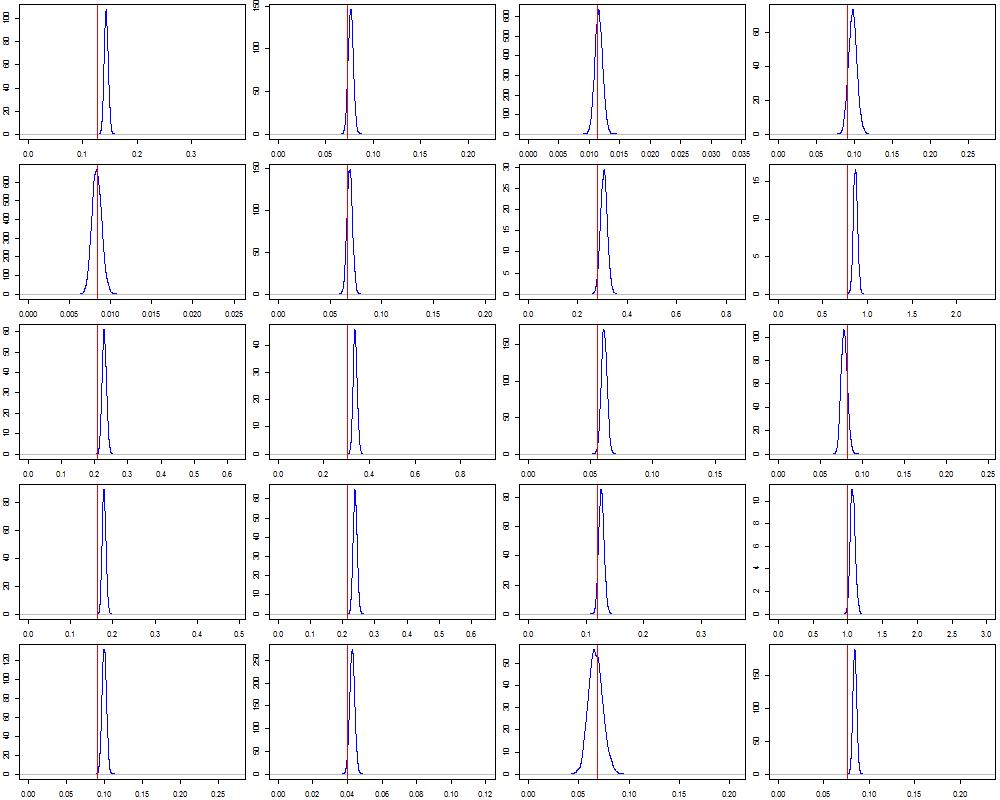}
    \caption{Posterior distributions of the habitat selection probabilities $S_{i2}$ for all $i$ ($S_{i1}=1$ for all $i$, each graph corresponds to a different value of $i$). The reference values are given in red.}
    \label{FigureDonneesSimuleesSpecificites}
    \end{center}
\end{figure}

{In this section we provide additional results to the ones presented in Section \ref{sectSimulateddata} for simulated data. Figure \ref{FigureDonneesSimuleesSpecificites}, as a complement to Figure \ref{FigureDonneesSimuleesAbondances}, provides the posterior distributions of the habitat selection probabilities $S_{i2}$ for all $i$, showing that these posterior are a good approximations to the reference values that we wish to estimate. Figure \ref{FigureBoxplotsDistSD}, as a complement to Figure \ref{FigureDonneesSimuleesAbondances} gives the boxplots of the mean of the squared difference between the posterior distribution and the real value of relative abundances, for the different considered models and datasets.}

\subsection{Additional results on real data}
\label{AppResultsRD}

\subsubsection{Some ecological results}

In this section we provide additional results with ecological motivations. 

\begin{figure}[ht]
\begin{center}
\includegraphics[scale=0.4, trim=0cm 0cm 0cm 0cm, clip]{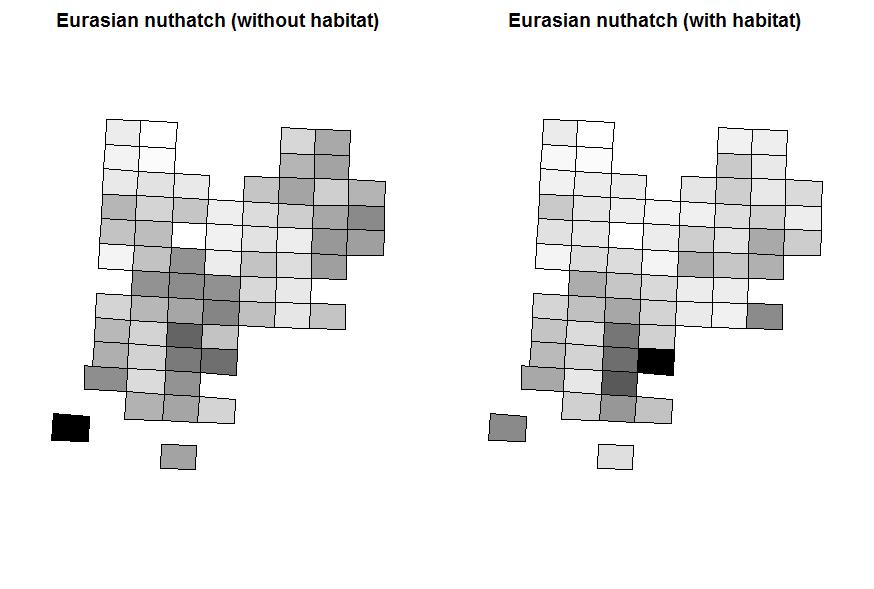}
    \caption{Relative density maps of the European nuthatch, without (left), and with (right) taking into account habitat structure. For each quadrat the gray level indicates the relative density $\frac{\hat N^{model}_{ij}V_1}{\hat N^{model}_{i1}V_j}$ where $V_l$ is the area of quadrat $l$.}
    \label{FigureMaps}
\end{center}
\end{figure}

\begin{figure}[ht]
\begin{center}
\begin{subfigure}{0.48\textwidth}
\begin{center}
\includegraphics[scale=0.15, trim=0cm 0cm 0cm 0cm]{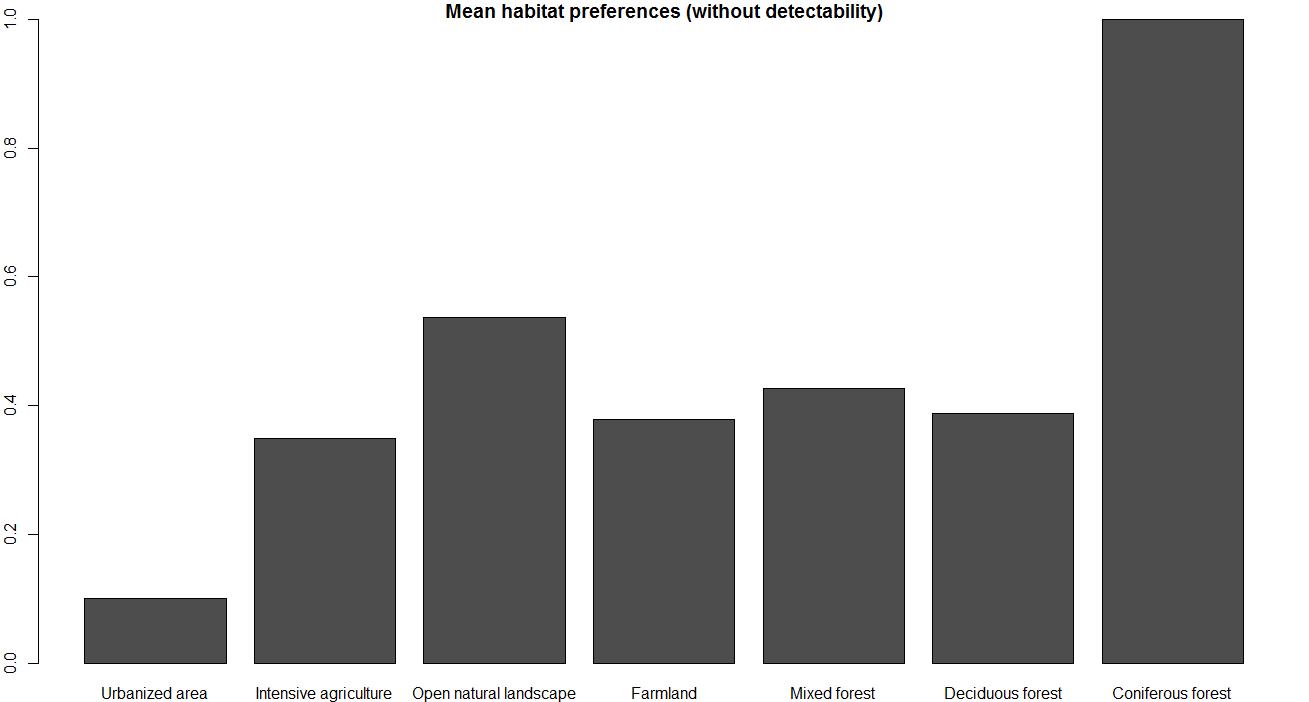}
    \end{center}
\end{subfigure}
\quad
\begin{subfigure}{0.48\textwidth}
\begin{center}
\includegraphics[scale=0.15, trim=0cm 0cm 0cm 0cm]{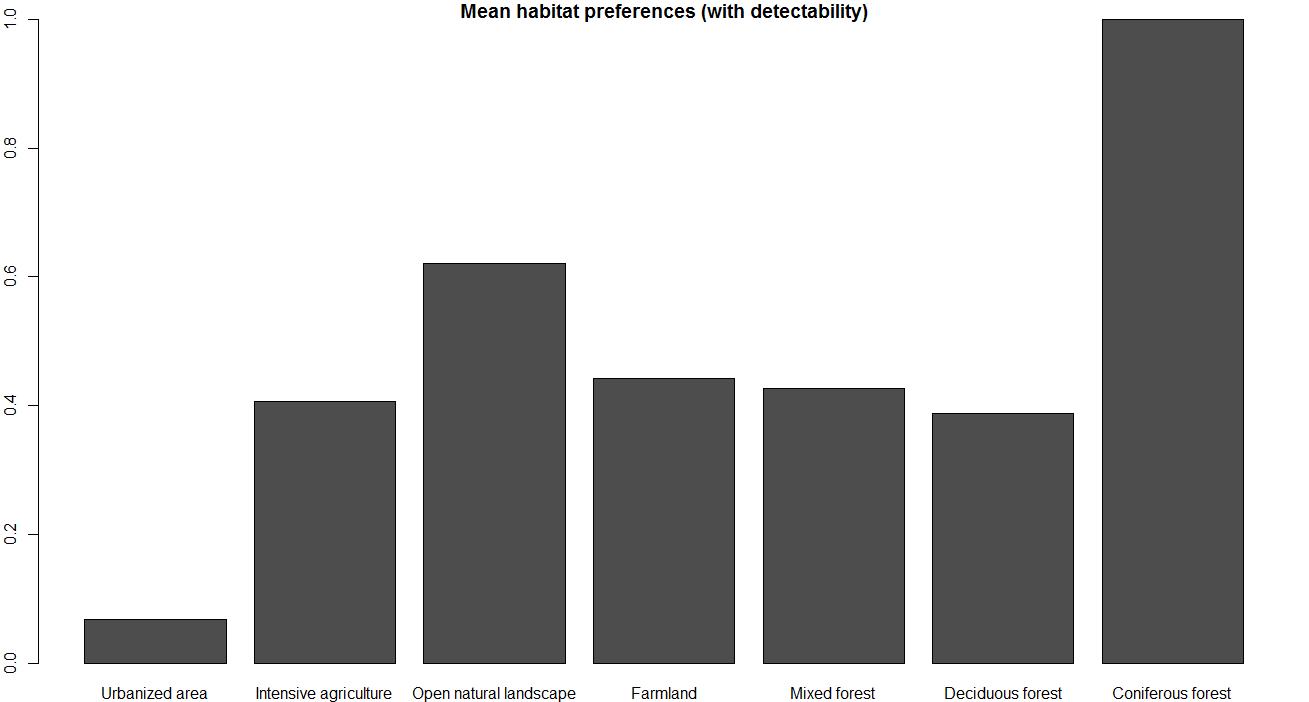}
\end{center}
\end{subfigure}
\caption{Mean species estimated habitat selection probabilities, without (left) and with (right) habitat dependent detectability.}
\label{FigurePreferences}
\end{center}
\end{figure}

 In Figure \ref{FigureMaps} we give a map of the estimated densities of the Grimpereau, with and without habitat struture. In Figure \ref{FigurePreferences} we give the mean preferences of all considered species, for each habitat type.

\subsubsection{Taking into account habitat dependent detectability}\label{sectHabitat}

We so far assumed that the detectability $P_{ik}$ of species $i$ in dataset $k$ does not depend on the habitat, which might be unrealistic since, in particular, the range of vision (or hearing) can be different from one habitat to the other. If so, our estimation of habitat selection parameters $S_{ih}$ but also of species relative abundances can be biased. Due to identifiability constraints, we cannot add and estimate an unknown list of parameters $\alpha_h$ taking into account the dependence of detectability to habitat (since they will be undistinguishable from the species habitat selection probabilities $S_{ih}$). Our proposition is to use an auxiliary dataset that can provide informations concerning the detectability associated to each considered habitat. We test this idea using a dataset provided by VigieNature that gives for different kinds of habitats the respective numbers of observations made in different ranges of distance. The program associated to this section is given in the file alpha.R. 

For each habitat $h$, we can assume that detection probability is equal to $1$ when observed individuals are "close enough" to the observer, since we only want to quantify the loss in detectability in each habitat due to the limitation in range of vision (or hearing) in this habitat. More precisely, we assume that detection probability is equal to $1$ when the observed individual is less than $25$ meters far from the observer. Then if we denote by $Y_h$ the number of observed individuals in habitat $h$ and by $Y_{1h}$ the number of observed individuals in habitat $h$, at distance less than $25$ meters from the observer, we can quantify the detectability in habitat $h$ by the quantity
$$\alpha_h=\frac{Y_{h}/Y_{1h}}{Y_{1}/Y_{11}}.$$
The result of these calculations is given in Table \ref{TableAlpha}. As expected, the detectability is lower in urbanized area and forest than in open and agricultural landscapes. The impact of taking into account habitat detectability is illustrated in Figure \ref{FigurePreferences}.

\begin{table}[ht]
\begin{center}
  \begin{tabular}{|c|c|}
    \hline
      Corine land Cover Habitat & Detectability $\alpha_h$\\
    \hline
      Urbanized area  & 1\\
     Intensive agriculture  & 1.72 \\
     Open natural landscape & 1.71\\
     Farmland with heterogenous landscape & 1.72\\
     Mixed forest & 1.47\\
     Deciduous forest & 1.47\\
     Coniferous forest & 1.47\\
    \hline
  \end{tabular}           
\caption{The detectability associated to each habitat, taking account differences in ranges of vision.} \label{TableAlpha}
\end{center}
\end{table}

\bibliographystyle{chicago}
\bibliography{mabiblio}

\end{document}